\documentstyle[aps,prl,twocolumn]{revtex}
\begin{document}
\title{Hydrogen-enhanced local plasticity in aluminum: an {\it ab initio} study} 
\author{Gang Lu}
\address
{Department of Physics, Harvard University, Cambridge, MA 02138}
\author{Qing Zhang, Nicholas Kioussis}
\address{Department of Physics, California State University Northridge,
Northridge, CA 91330}
\author{Efthimios Kaxiras}
\address{Department of Physics, 
Harvard University, Cambridge, MA 02138}
\address{
\begin{center} 
\begin{minipage}{14cm}
\begin{abstract}
Dislocation core properties of Al with and without H impurities are studied 
using the Peierls-Nabarro model with parameters determined 
by {\it ab initio} calculations.
We find that H not only facilitates dislocation emission from the
crack tip but also enhances dislocation mobility dramatically, leading to
macroscopically softening and thinning of the material ahead of the crack tip.
We observe strong
binding between H and dislocation cores, with the binding energy depending on 
dislocation character. 
This dependence can directly affect the mechanical 
properties of Al by inhibiting dislocation cross-slip and developing slip
planarity.  
\end{abstract}
\end{minipage}
\end{center}
}
\maketitle

The mechanical properties of solids are not only a function of their intrinsic
atomic structure, but also of the environment in which they exist. Small amounts
of impurity atoms in the environment can drastically change the response of a solid 
to external loading.
Hydrogen embrittlement is one of the most important and 
well-studied processes of environmental degradation of materials. 
After over a century of studies, the definitive mechanism of H 
embrittlement still remains unknown although significant progress has been made
toward developing a detailed understanding of the problem.
Three general mechanisms for H embrittement have been proposed:
(1) Formation of a hydride phase; 
(2) H-enhanced local plasticity; and (3) H-induced grain boundary weakening \cite{RMP}. 
The underlying atomic 
processes and the relative importance of the three mechanisms, however, remain uncertain
and controversial. In this paper, 
we study a key aspect of H embrittlement, referred to as H-enhanced local plasticity
(HELP) \cite{Beachem}, which has attracted much attention recently. 
There seems to be overwhelming evidence that plasticity is a fundamental
contributor to H embrittlement of virtually all susceptible metals and alloys, 
including Al, the model material we consider in this paper
\cite{RMP,Ref,Zeides}. The most dramatic experimental observation  
is that H increases  
dislocation mobility significantly under constant
stress. This H-enhanced mobility is observed for screw, edge and mixed dislocations
as well as for isolated dislocations and dislocation tangles \cite{RMP,Sofronis}. 
Other important experimental results 
include the observations of slip planarity \cite{Zeides,Ref2} and 
strong binding between H and dislocation cores \cite{RMP}. 

In contrast to the vast body of experimental evidence, theoretical studies 
of the HELP mechanism are scarce.
To our knowledge, the only theoretical investigation of the HELP mechanism  
is based on finite-element modeling of 
elastic interactions among dislocations in the presence of H \cite{Sirois}. 
Although this approach was
able to explain the H-enhanced dislocation mobility through the H-induced ``elastic 
shield'' effect, it completely ignores the effects of H on dislocation core structure 
and contains no information for the relevant atomic processes, which are critical for a 
complete understanding of the phenomenon.  
The purpose of our work is to provide a 
comprehensive 
understanding of the aforementioned experimental results based on an  
{\it ab initio} theoretical framework, which hopefully can set the stage for an 
atomistic theory of H embrittlement.  
             
In this paper, we have used the recently developed Semidiscrete Variational 
Peierls-Nabarro (SVPN)
model \cite{Bulatov,Lu,Lu1} with {\it ab initio} determined  
$\gamma$-surfaces \cite{GSF} and 
elastic constants. 
This approach has been shown 
to predict reliable dislocation core properties by comparing 
its predictions to direct atomistic 
simulations based on the same force law as that used for the $\gamma$-surface
calculations \cite{Bulatov,Lu}. The uniqueness of
the approach when combined with {\it ab initio} calculations for the energetics is 
that it produces essentially an atomistic simulation for dislocation 
core properties without suffering from the uncertainties associated with
empirical potentials. For example, atomistic simulations based on 
EAM potentials predict that dislocations in Al will dissociate into partials \cite
{Bulatov2,Mills}, while experimentally no such dissociation is observed \cite{Duesbery},
a discrepancy which can be traced to the fact
the EAM potentials underestimate the intrinsic stacking 
fault energy \cite{Lu}.  

In the SVPN approach, the equilibrium structure of a dislocation 
is obtained by minimizing the dislocation energy functional
\begin{equation}
U_{disl} = U_{elastic} + U_{misfit} + U_{stress} + Kb^2{\rm ln}L,
\end{equation}
with respect to the dislocation displacement density \cite{Bulatov,Lu,Lu1}. 
The $\gamma$-surface determined 
from {\it ab initio} calculations enters the equation through the $U_{misfit}$ term. 
We identify the dislocation configuration-dependent part of the elastic energy and
the misfit energy as the core energy, $U_{core} = U_{elastic} + U_{misfit}$,
while the configuration-independent part of the elastic energy $Kb^2{\rm ln}L$ is
excluded because it is irrelevant in the variational procedure and it has no contribution
to the dislocation core structure. 
The response of a dislocation to an applied stress is achieved by the minimization of
the energy functional at
the given value of the applied stress.
An instability is reached when an optimal solution for the density distribution no longer
 exists, which
is manifested numerically by the failure of the minimization procedure to convergence.
The Peierls stress is then defined as the critical value of the applied stress
giving rise to this instability.

In order to study the interaction between H and dislocations, we have carried out 
{\it ab initio}
calculations of the $\gamma$-surface with H placed at interstitial sites of the Al 
lattice. 
Specifically, we have used a large supercell containing six Al layers in the [111]
direction with 
four atoms per layer to simulate a low concentration of H impurity ($\sim$ 4 at.\%). 
Two sets 
of calculations were performed with H placed at the octahedral and the 
tetrahedral sites in Al. The {\it ab initio} calculations are 
based on the pseudopotential \cite{Vanderbilt} plane-wave method with the generalized 
gradient approximation to the exchange-correlation functional \cite{GGA}. A kinetic 
energy cutoff of 12 Ry for the plane-wave basis is used and a $k$-point grid consisting 
of (8,8,4) divisions along the reciprocal lattice vectors is sampled for the Brillouin
zone integration. In our calculations, the H atom is allowed to relax freely 
according to the
Hellmann-Feynman force acting on it while Al atoms are restrained to move only 
along the [111] direction.  
Volume relaxation is performed for each sliding distance to minimize the tensile stress 
on the supercell. 
Table I lists the various stacking fault energies obtained for the pure Al and 
the Al+H systems.
Although the octahedral site is more stable for H without 
sliding, the energy for the tetrahedral site can be lower for certain sliding vectors. 
Therefore the energies in Table I and Fig. 1 are given with respect to the lowest  
values of
the two sites for each sliding. From Table I we find that H decreases the
energy of the various stacking faults by up to 50\%, which could change dislocation 
properties significantly. For example,  
one can predict that dislocations would be emitted 
more easily from a crack tip when H is present based on the lowering of the 
unstable stacking
fault energy along the [1$\bar{2}$1] direction \cite{Rice}. In fact, this prediction is in 
accordance  
with atomistic simulations for Ni, where dislocations are found to be emitted more 
rapidly from a crack 
tip in the presence of H \cite{Daw}. 

The $\gamma$-surface for both the pure Al and the Al+H systems 
is shown in Fig. 1(a) and (b) 
respectively. The energy along various high symmetry directions is determined from 
{\it ab initio} calculations while the rest of the $\gamma$-surface is obtained 
by fitting  
a symmetrized polynomial basis, which guarantees the correct rotational symmetry of the
 $\gamma$-surface on the (111) plane. Comparing the two $\gamma$-surfaces, 
we find an 
overall reduction in energy in the presence of H. In order to understand how
 H affects the elastic properties of Al, we also calculate the various 
elastic constants 
for both the pure Al and the Al+H systems using {\it ab initio} approaches.  
We used a smaller
unit cell with only four Al atoms and one H impurity to facilitate these calculations, 
and find that, even at such a high concentration of H, the prelogarithmic 
elastic energy factors entering the model \cite{Lu} change only by a few percent at most
\cite{elast}. 

Having determined all the necessary parameters that enter the model, we can start to
investigate the core properties for various dislocations in Al. We have 
studied screw (0$^\circ$), 30$^\circ$, 60$^\circ$ and edge (90$^\circ$)
dislocations. 
The Peierls stress for these dislocations is shown in Table II. We find that
the Peierls stress, which is the minimum stress to move a stationary dislocation, 
is reduced by more than an
order of magnitude in the presence of H. 
Moreover this H-enhanced mobility is observed for 
screw, edge and mixed dislocations, which is compatible with the experimental results.
Not only is our calculation the first successful attempt to explain HELP associated
with individual dislocation behavior, it also invalidates the perception that the 
only plausible explanation of HELP must be based on elastic interactions among 
dislocations and that dislocation-lattice interaction is not important \cite{RMP}. On the 
contrary, we find the dislocation-lattice interaction to be very important and 
that the dislocation
core structure is responsible for 
the observed  H-enhanced dislocation mobility.  
Our result, however, does not exclude the possibility that 
dislocation-dislocation interactions do play a role in HELP.  

Other important experimental results that any theory of H embrittlement must explain 
are the observed H trapping 
in dislocation cores and the H-induced slip planarity. To this end, we have calculated  
the energetics of dislocations for the pure Al and the Al+H systems. The results are 
shown in Table II. We remind the reader that $E_{core}$ listed here is only
the displacement density dependent part of the total energy, excluding the large 
positive elastic
term $Kb^2{\rm ln}L$ which renders the total energy positive \cite{Lu}. 
The binding energy of H is defined as the difference between 
dislocation core energies with and without the presence of H. 
From these results 
we find that there is strong binding between H and dislocation cores and 
that H is attracted (trapped) to dislocation cores to 
lower the core energies. More importantly, we find that the binding energy is 
a function
of dislocation character, with edge dislocations having the greatest and screw 
dislocations
having the lowest binding energy, respectively. 
For a mixed dislocation, we find that the binding energy increases with the amount of  
edge component. These results have significant consequences for
the experimentally observed inhibition of cross-slip 
and slip planarity in the presence of H. Experiments show that in the 
presence of H slip is confined to the primary slip plane and dislocation cross-slip is
restricted, which leads to the observed slip planarity. 
On the other hand, removal of H from 
the sample allows a dislocation segment on the primary plane to reorient towards the
screw orientation and to complete the cross-slip precess \cite{Ferreira}. It was believed 
from the experimental evidence that any process, such as cross-slip, that increases the 
screw component length at the expense of the edge component will be made more difficult 
by the presence of H \cite{Ferreira}. Our results for the binding energy 
of H in dislocation cores are 
the key for understanding these experimental observations: 
Since the edge dislocation has twice the binding energy as the screw,  
it will cost much more energy for an edge dislocation in the presence of H 
to transform to a screw 
dislocation in order to cross-slip.
In the same vein, it costs more energy for a mixed dislocation 
to transfer its edge component to a screw dislocation for cross-slip.
In other words, H can stabilize the edge component of a mixed dislocation and 
inhibit the dislocation
cross-slip process, leading to the observed slip planarity.
It is also interesting to point out that our  
results for the binding energy of H are in qualitative agreement with those 
from atomistic simulations of H  
in Ni, where the binding energies of H to edge dislocations and screw dislocations are
found to be 0.12 eV and 0.09 eV per H atom, respectively \cite{Angelo}. 

To further understand these results, we calculate the displacement density for the 
various
dislocations which is presented in Fig. 2. We find that although the 
intrinsic stacking fault energy is 
reduced by almost 40\% in the presence of H, dislocations are not dissociated into 
partials \cite{partial}. 
This result is important because it discredits the argument that the observed 
restriction of 
cross-slip may result from the increased separation of partials due to the 
reduction of the stacking fault energy \cite{Ref3}. In fact, our result is  
consistent with the experimental findings that no widely spaced dislocation 
partials were ever observed
in TEM studies of Al containing large concentrations of H \cite{Ferreira}. Therefore,
we believe that the observed slip planarity and restriction of cross-slip is due to the 
strong binding between H and the edge component of dislocations, not only for Al, but 
possibly also
for other metals \cite{Fer2,Zeides,Ferreira}.

Based on the experimental observations and the results we obtained in this study, 
we identify the following processes which we believe could lead to the H-induced 
fracture: 
(a) {\em H is attracted to crack tips}: Application of external stress produces 
local concentration 
of tensile stress in the vicinity of cracks, which attracts H 
since H prefers to stay
in slightly enlarged interstitial sites. (b) {\em Thinning and softening}:
 These segregated H impurities at the 
crack tip can facilitate dislocation generation and enhance dislocation mobility, 
which will lead to extensive plastic deformation in front of the crack,
causing thinning and softening of the material ahead the crack. (c) {\em Lowering fracture
stress}: The thinning and 
softening processes, along with the localization of slip due to the inhibition of 
cross-slip,
allow the crack to propagate 
at lower stress levels, prior to general yielding away from the crack tip. 
Although our calculations concern Al, we believe that the results are also applicable
to other metals. Furthermore we
should point out that the above processes
are not likely to be the only mechanism operating in
H embrittlement \cite{Sofronis}, but are the most relevant processes 
based on HELP theory.

In summary, we have studied the interaction of interstitial H with dislocations
in Al using the SVPN model with {\it ab initio}
determined parameters.  
We find that H can lower the Peierls stress of  
dislocations by more than an order of magnitude, strongly supporting the HELP model.
We also find there is a strong binding 
of H to dislocation cores and that the binding energy is a function of dislocation
character.  
Our findings
explain the experimentally observed restriction of the cross-slip process 
and slip planarity in the presence of H. The dislocations do not 
dissociate into partials even though the intrinsic stacking fault energy is reduced
by 40\% in the presence of H. Finally, based on the results of our calculations, we
 identify the processes which could lead to
H embrittlement.  

The work at California State University Northridge was supported by Grant No.
DAAG55-97-1-0093 from the US Army Research Office. The work at Harvard University 
was supported by Grant No. F49620-99-1-0272 from the US Air Force Office for 
Scientific Research. We would like to thank Daniel 
Orlikowski for useful discussions.

\begin{figure}
\caption{The $\gamma$-surfaces (J/$m^2$) for displacements along a (111) plane
for (a) pure Al and (b) Al+H systems.
The corners of the plane and its center correspond to
identical equilibrium configurations,
i.e., the ideal lattice.
The two energy surfaces are displayed
in exactly the same perspective and on the
same energy scale to facilitate comparison of important features.}
\label{fig1}
\end{figure}

\begin{figure}
\caption{Dislocation displacement density for four dislocations (clockwise): 
screw (0$^\circ$), 30$^\circ$, 60$^\circ$ and edge (90$^\circ$) for the pure Al (solid lines) 
and the Al+H (dashed lines) systems.} 
\end{figure}

\begin{table}
\caption{Fault vectors and energies (J/m$^2$) for four important stacking faults
 of the pure Al and the Al+H systems.} 
\begin{tabular}{cccc}
                    & Vector & Al  &  Al+H \\ \hline
 Intrinsic stacking & 1/6[12$\bar{1}$]  & 0.164 & 0.103 \\
 Unstable stacking  & 1/10[12$\bar{1}$] & 0.224 & 0.126 \\
 Unstable stacking  & 1/4[101]          & 0.250 & 0.132 \\
 Run-on stacking    & 1/3[12$\bar{1}$]  & 0.400 & 0.310\\
\end{tabular}
\end{table}

\begin{table}
\caption{Peierls stress ($\sigma_p$, MPa), core energies ($E_{core}$, eV/\AA) 
for the four dislocations 
in the pure Al and the Al+H systems and
binding energy ($E_{b}$, eV/atom) for the four dislocations.}
\begin{tabular}{cccccc}

  &       & screw &  30$^\circ$  & 60$^\circ$ & edge \\ \hline
$\sigma_p$ & Al & 254  & 51 & 97 & 3 \\ \cline{2-6}
           & Al+H & 1.7 & 1.2 & 1.4 & 0.3 \\ \hline
$E_{core}$ &  Al& -0.08 & -0.11 &     -0.17       & -0.20 \\ 
\cline{2-6}
           & Al+H&-0.14 & -0.18 & -0.27 & -0.32 \\ \hline
$E_{b}$    &     &0.06 & 0.07  & 0.10 & 0.12 \\
\end{tabular}
\end{table}   


\begin{thebibliography}{99}
\bibitem{GSF}
When a crystal is cut along its slip plane and the upper half is displaced
relative to the lower by a vector $\vec{f}$, the energy increase per unit area
is defined as $\gamma$ and the energy surface 
$\gamma(\vec{f})$, with $\vec{f}$ spanning the entire slip plane, is 
called the $\gamma$-surface.
\bibitem{elast}
The prelogarithmic elastic energy factors $K_e$ and $K_s$ for Al (Al+H) are 
7.00 (7.13) GPa and 4.59 (4.68) GPa, respectively.
\bibitem{partial}
The small double-peak in the 60$^\circ$ dislocation does not suggest dissociation.
It actually comes from the non-equivalent atomic spacing of the lattice (see Ref. 9).
\end{thebibliography}
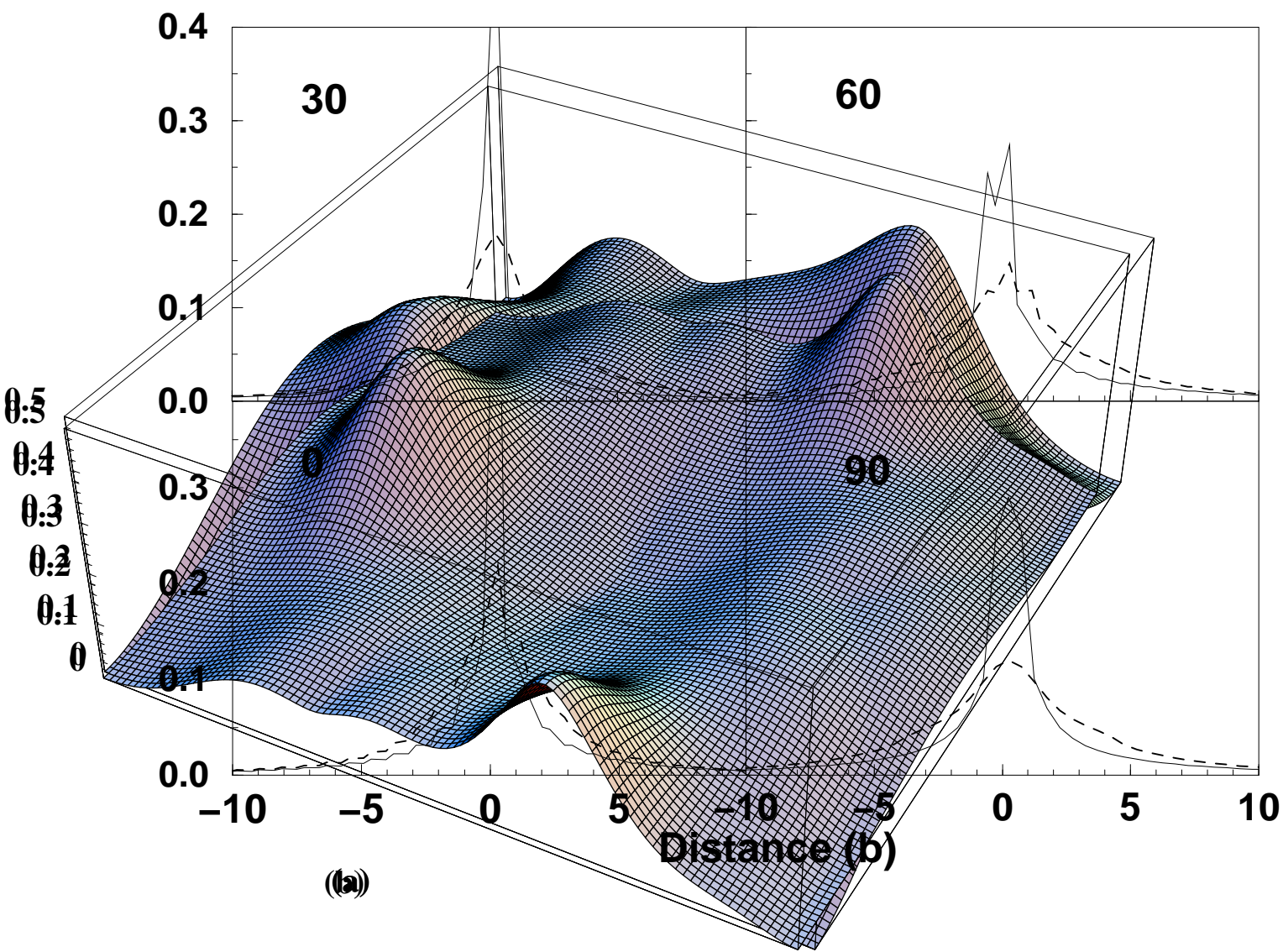
\end{document}